\documentstyle[preprint,prb,aps]{revtex}

\begin{document}

\title{Measurement of Resonant Frequency and Quality 
Factor of Microwave Resonators:\\
Comparison of Methods}

\author{Paul J. Petersan and Steven M. Anlage}

\address{Center for Superconductivity Research, 
Department of Physics, University of\\
Maryland, College Park, MD 20742-4111 }

\maketitle

\begin{abstract}
Precise microwave measurements of sample conductivity, dielectric, and
magnetic properties are routinely performed with cavity perturbation
measurements. These methods require the accurate determination of quality
factor and resonant frequency of microwave resonators. Seven different
methods to determine the resonant frequency and quality factor from complex
transmission coefficient data are discussed and compared to find which is
most accurate and precise when tested using identical data. We find that the
nonlinear least-squares fit to the phase vs. frequency is the most accurate
and precise when the signal-to-noise ratio is greater than 65. For noisier
data, the nonlinear least squares fit to a Lorentzian curve is more accurate
and precise. The results are general and can be applied to the analysis of
many kinds of resonant phenomena.
\end{abstract}


\section{Introduction}

Our objective is to accurately and precisely measure the quality factor $Q$,
and resonant frequency $f_o,$ of a microwave resonator, using complex
transmission coefficient data as a function of frequency. Accurate $Q$ and 
$f_o$ measurements are needed for high precision cavity perturbation
measurements of surface impedance, dielectric constant, magnetic
permeability, etc. Under realistic experimental conditions, corruption of
the data occurs because of cross-talk between the transmission lines and
between coupling structures, the separation between the coupling ports and
measurement device, and noise. Although there are many methods discussed in
the literature for measuring $Q$ and resonant frequency, we are aware of no
treatment of these different methods which quantitatively compares their
accuracy or precision under real measurement conditions. In practice, the $Q$
can vary from 10$^7$ to 10$^3$ in superconducting cavity perturbation
experiments, so that a $Q$ determination must be robust over many orders of
magnitude of $Q$. Also, it must be possible to accurately determine $Q$ and 
$f_o$ in the presence of modest amounts of noise. In this paper we will
determine the best methods of evaluating complex transmission coefficient
data, i.e. the most precise, accurate, robust in $Q$, and robust in the
presence of noise.

Many different methods have been introduced to measure the quality factor
and resonant frequency of microwave cavities over the past fifty years.
Smith chart methods have been used to determine half power points which can
be used in conjunction with the value of the resonant frequency to deduce
the quality factor of the cavity.\cite{Montgomery}$^{-}$\cite{Sun} In the
decay method for determining the quality factor, the fields in the cavity
are allowed to build up to equilibrium, the input power is turned off, and
the exponential decrease in the power leaving the cavity is measured and fit
to determine the quality factor of 
the cavity.\cite{Ginzton,Sucher,Barmatz,Padamsee} 

Cavity stabilization methods put the
cavity in a feedback loop to stabilize an oscillator at the resonant
frequency of the cavity.\cite{Padamsee}$^{-}$\cite{Luiten} For one port
cavities, reflection measurements provide a determination of the half power
points and also determine the coupling constant, allowing one to calculate
the unloaded $Q$.\cite{Aitker}$^{-}$\cite{Subramanian} In more recent years,
complex transmission coefficient data vs. frequency is found from vector
measurements of transmitted signals through the 
cavity.\cite{Sanchez}$^{-}$\cite{Leong} 
Methods which use this type of data to determine $Q$ and $f_o$ are the subject of this paper.

We have selected seven different methods for determining $f_o$ and $Q$ from
complex transmission coefficient data. We have collected sets of 'typical'
data from realistic measurement situations to test all of the $Q$ and $f_o$
determination methods. We have also created data and added noise to it to
measure the accuracy of the methods. In this paper we consider only random
errors and not systematic errors, such as vibrations of the cavity which
artificially broaden the resonance.\cite{Padamsee}$^{-}$\cite{Luiten} After
comparing all of the different methods, we find that the nonlinear least
squares fit to the phase vs. frequency and the nonlinear least squares fit
of the magnitude of the transmission coefficient to the Lorentzian curve are
the best methods for determining the resonant frequency and quality factor.
The phase vs. frequency fit is the most precise and accurate over many
decades of $Q$ values if the signal-to-noise ratio (SNR) is high 
(SNR \mbox{$>$} 65), however the Lorentzian fit is more robust for noisier data. Some of the
methods discussed here rely on a circle fit to the complex transmission
coefficient data as a step to finding $f_o$ and $Q$. We find that by
adjusting this fitting we can improve the determination of the quality
factor and resonant frequency, particularly for noisy data.

In section II of this paper, the simple lumped element model for a microwave
resonator is reviewed and developed. A description of our particular
experimental setup is then given, although the results of this paper apply
to any transmission resonator. We then discuss the data collected and
generated for use in the method comparison in section III. Section IV
outlines all of the methods that are studied in this paper. It should be
noted that each method is tested using exactly the same data. The results of
the comparison are presented and discussed in section V. Possible
improvements for some of the methods follow in section VI, and the
concluding remarks of the paper are made in the final section.

\section{Lumped Element Model of a Resonator}

To set the stage for our discussion of the different methods of determining 

$Q$ and resonant frequency, we briefly review the simple lumped-element model
of an electromagnetic resonator. As a model for an ideal resonator, we use
the series RLC circuit (see inset of Fig. \ref{Lorentzian}), defining 
$1/2\pi \sqrt{LC}$ as the resonant frequency $f_o$.\cite{Ma} The quality
factor is defined as 2$\pi $ times the ratio of the total energy stored in
the resonator to the energy dissipated per cycle.\cite{Sucher} For the
lumped element model in Fig. \ref{Lorentzian}, the quality factor $Q$ is 
$2\pi f_oL/R$. The resonator is coupled to transmission lines of impedance 
$Z_o$ by the mutual inductances $l_{m1}$ and $l_{m2}$. The complex
transmission coefficient, $S_{21}$ (ratio of the voltage transmitted to the
incident voltage), as a function of driving frequency $f$, is given in the
limit of weak coupling by:\cite{Ma}
\begin{equation}
S_{21}\left( f\right) =\frac{\overline{S_{21}}}{1+iQ\left( \frac f{f_o}-
\frac{f_o}f\right) }  
\label{0}
\end{equation}
The additional assumption that $f$ $\symbol{126}$ $f_o$ near resonance
simplifies the frequency dependence in the denominator resulting in:
\begin{equation}
S_{21}\left( f\right) =\frac{\overline{S_{21}}}{1+i2Q\left( \frac f{f_o}
-1\right) }  
\label{1}
\end{equation}
where $\overline{S_{21}}$ is the maximum of the transmission coefficient
which occurs at the peak of the resonance:
\begin{equation}
\overline{S_{21}}=\frac{8\pi ^2f^2l_{m1}l_{m2}}
{Z_oR}=2\sqrt{\beta _1\beta _2}  
\label{2}
\end{equation}
Here $R$ is the resistance in the circuit model and this expression again is
valid in the weak coupling limit. On the far right side of Eq. (\ref{2}), 
$\beta _1$ and $\beta _2$ are the coupling coefficients on ports 1 and 2,
respectively,\cite{Ginzton}$^{,}$\cite{Leong} where 
$\beta_j=\left( 2\pi f\right) ^2l_{mj}^2/Z_oR$, with $j=1,2$.
The magnitude of the complex transmission coefficient is:
\begin{equation}
\left| S_{21}\left( f\right) \right| =\frac{\left| \overline{S_{21}}\right| 
}{\sqrt{1+4Q^2\left( \frac f{f_o}-1\right) ^2}}
\label{3}
\end{equation}
The plot of $\left| S_{21}\right| $ vs. frequency forms a Lorentzian curve
with the resonant frequency located at the position of the maximum magnitude
(Fig. \ref{Lorentzian}). A numerical investigation of $\left| S_{21}\right| $
with and without the simplified denominator assumption leading to 
Eq. (\ref{1}), shows that even for a relatively low $Q$ ($Q=$100), 
the difference between the magnitudes is 
less than half a percent of the magnitude using
Eq. (\ref{0}). For larger Q the difference is much smaller, so we take this
assumption as valid. All of the analysis methods treated in this paper make
use of the simplified denominator assumption, as well as all the data we
create to test the methods.

The plot of the imaginary part of $S_{21}$ (Eq. (\ref{1}))versus the real
part (with frequency as a parameter), forms a circle in canonical position
with its center on the real axis (Fig. \ref{Circles}). The circle intersects
the real axis at two points, at the origin and at the location of the
resonant frequency.

Important alterations to the data occur when we take into account several
aspects of the real measurement situation. The first modification arises
when considering the cross talk between the cables and/or the coupling
structures. This introduces a complex translation 
$X=\left( x_o,y_o\right)$, of the center of the circle away from its place on the real axis.\cite{Ma}$^{,}$\cite{Leong}$^{,}$\cite{Snortland} Secondly, a phase shift $\phi $ is
introduced because the coupling ports of the resonator do not necessarily
coincide with the plane of the measurement. This effect rotates the circle
around the origin 
(Fig. \ref{Circles}).\cite{Ma}$^{,}$\cite{Leong}$^{,}$\cite{Snortland} 
The corrected complex transmission coefficient, 
$\widetilde{S}_{21}$, is then given by:
\begin{equation}
\widetilde{S}_{21}=\left( S_{21}+X\right) e^{i\phi }  
\label{4}
\end{equation}
It should be noted that the order in which the translation and rotation are
performed is unique.\cite{Snortland}

Any method of determining $Q$ and $f_o$ from complex transmission data must
effectively deal with the corruption of the data represented by Eq. (\ref{4}). In addition, the method used to determine $f_o$ and $Q$ must give
accurate and precise results even in the presence of noise. This is
necessary since, in typical measurements, $Q$ ranges over several orders of
magnitude causing the signal-to-noise ratio (SNR, defined in section III.
c.) during a single data run to vary significantly. Further corruption of
the data can occur if there are nearby resonances present, particularly
those with lower $Q$. This introduces a background variation onto the
circles shown in Fig. \ref{Circles} and may interfere with the determination
of $f_o$ and $Q$. In this paper we consider only single isolated resonances
and refer the reader to an existing treatment of multiple 
resonances.\cite{Gao}

\section{Data Used for Method Comparison}

In this section we discuss the data we use for making quantitative
comparisons of each method. The data is selected to be representative of
that encountered in real measurement situations. Each trace consists of 801
frequency points, each of which have an associated real and imaginary part
of $S_{21}$. Two types of data have been used for comparing the methods;
measured data and generated data. The measured data is collected with the
network analyzer and cavity described below. The generated data is
constructed to look like the measured data, but the underlying $Q$ and
resonant frequency are known exactly. All of the methods discussed in the
next section are tested using exactly the same data.

\subsection{Measured Data}

Complex transmission coefficient vs. frequency data is collected using a
superconducting cylindrical Niobium cavity submerged in liquid Helium at 4.2
K. Microwave coupling to the cavity is achieved using magnetic loops located
at the end of 0.086'' coaxial cables. The loops are introduced into the
cavity with controllable position and orientation. The coaxial cables come
out of the cryogenic dewar and are then connected to an HP8510C vector
network analyzer.\cite{Mao} The cavity design\cite{Sridhar} has recently
been modified to allow top-loading of the samples into the cavity.

A sample is introduced into the center of the cavity on the end of a

sapphire rod. The temperature of the sample can be varied by heating the
rod, with a minimal perturbation to the superconducting Nb walls. The
quality factor of the cavity resonator in the TE$_{011}$ mode can range from
about 2 $\times $ 10$^7$ to 1 $\times $ 10$^3$, with a resonant frequency of
approximately 9.6 GHz. In a typical run with a superconducting crystal,
where the temperature varies from 4.2 K to 200 K, $f_o$ decreases by about
10 MHz and $Q$ changes from about 1 $\times $ 10$^7$ to 4 $\times $ 10$^3$.
For accurate measurement of the electrodynamic properties of samples, it is
important to be able to resolve frequency shifts of the cavity as small as 1
Hz at low temperatures.

\subsubsection{Fixed Powers}

One hundred $S_{21}$ vs. frequency traces were taken using the network
analyzer held at a fixed power and with constant coupling to the cavity. One
such data set was made with the source power at $+$15 dBm (SNR $\approx $
368, $f_o\approx $ 9.600242 GHz, $Q$ $\approx $ 6.39 $\times $ 10$^6$),
another set was taken with the source power at $+$10 dBm (SNR $\approx $
108, $f_o\approx $ 9.599754 GHz, $Q$ $\approx $ 6.46 $\times $ 10$^6$), a
third data set was taken with the source power at $+$3 dBm (SNR $\approx $
49, $f_o\approx $ 9.599754 GHz, $Q$ $\approx $ 6.50 $\times $ 10$^6$). (The
approximate values for $f_o$ and $Q$ are obtained from the phase vs.
frequency averages discussed below)

\subsubsection{Power Ramp}

To collect data with a systematic variation of signal-to-noise ratio, we
took single traces at a series of different input powers. A power-ramped
data set was taken in a cavity where controllable parameters, such as
temperature and coupling, were fixed, the only thing that changed was the
microwave power input to the cavity. An $S_{21}$ vs. frequency trace was
taken for powers ranging from $-$18 dBm to $+$15 dBm, in steps of 0.5 dBm.
This corresponds to a change in the signal-to-noise ratio from about 5 to
168 ($f_o\approx $ 9.603938 GHz, $Q$ $\approx $ 8.71 $\times $ 10$^6$).

\subsection{Generated Data}

To check the accuracy of all the methods, we generated data with known
characteristics, and added a controlled amount of noise to simulate the
measured data. The data was created using the real and imaginary parts of an
ideal $S_{21}$ as a function of frequency Eq. (\ref{1}); 
\begin{equation}
\mathop{\rm Re}
S_{21}\left( f\right) =\frac{\overline{S_{21}}}{1+4Q^2\left( \frac f{f_o}
-1\right) ^2}\quad \quad \hspace{0in}
\mathop{\rm Im}
S_{21}\left( f\right) =\frac{-\overline{S_{21}}2Q\left( \frac f{f_o}
-1\right) }{1+4Q^2\left( \frac f{f_o}-1\right) ^2}  
\label{5}
\end{equation}
Where $\overline{S_{21}}$ is the diameter of the circle being generated (see
Fig. \ref{Circles}), $Q$ is the quality factor, and $f_o$ is the resonant
frequency, which are all fixed. The frequency $f$, is incremented around the
resonant frequency to create the circle. There are 400 equally spaced
frequency points before and after the resonant frequency, totaling 801 data
points. The total span of the generated data is about four 3dB bandwidths
for all Q values.

To simulate measured data, noise was added to the data using Gaussian
distributed random numbers\cite{Press} that were scaled to be a fixed
fraction of the radius, $r$ of the circle described by the data in the
complex $S_{21}$ plane. The noisy data was then translated and rotated to
mimic the effect of cross talk in the cables and coupling structures, and
delay (Eq. (\ref{4})).

\subsubsection{Power Ramp}

A power ramp was simulated by varying the amplitude of the noise added to
the circles. A total of 78 $S_{21}$ vs. frequency traces were created with a
variation of the signal-to-noise ratio from about 1 to 2000 ($f_o$ $=$ 9.600
GHz, $Q$ $=$ 1.00 $\times $ 10$^6$, $x_o=$ 0.1972,$\ y_o=-$0.0877, $r=$ 0.2, 
$\phi =\pi $/17)

\subsubsection{Fixed $Q$ Values}

Data with different fixed $Q$ values were created using the above real and
imaginary expressions for $S_{21}$. Groups of data were created with 100
traces each using: $Q$ = 10$^2$, 10$^3$, 10$^4$, 10$^5$ ($f_o$ = 9.600 GHz
and SNR $\approx $ 65 for all sets). They include fixed noise amplitude, and
were each rotated and translated equal amounts to simulate measured data. 
($x_o=$ 0.01, $y_o=$ 0.015, $r=$ 0.2, $\phi =\pi $/19)

\subsection{Signal-to-Noise Ratio}

The signal-to-noise ratio was found for all data sets by first determining
the radius $r_{circle}$, and center $(x_c,y_c)$ of the circle when plotting
the imaginary part of the complex transmission coefficient vs. the real part
(Fig. \ref{Circles}). Next, the distance to each 
data point $\left(x_i,y_i\right) $ ($i=$ 1 to 801) 
from the center is calculated from: 
\begin{equation}
d_i=\sqrt{\left( x_i-x_c\right) ^2+(y_i-y_c)^2}
\end{equation}
The signal-to-noise ratio is defined as: 
\begin{equation}
SNR=\frac{r_{circle}}{\sqrt{\frac 1{800}\sum\limits_{i=1}^{801}\left(
d_i-r_{circle}\right) ^2}}
\end{equation}
In the case of generated data, where the center and radius of the circle are
known, the SNR is very well defined. However, the SNR values are approximate
for the measured data because of uncertainties in the determination of the
center and radius of the circles.

\section{Description of Methods}

In this section we summarize the basic principles of the leading methods for
determining the $Q$ and resonant frequency from complex transmission
coefficient vs. frequency data. Further details on implementing these
particular methods can be found in the cited references. Because we believe
that this is the first published description of the inverse mapping
technique, we shall discuss it in more detail than the other methods. The
Resonance Curve Area and Snortland techniques are not widely known, hence a
brief review of these methods is also included.

The first three methods take the data as it appears and determine the $Q$
from the estimated bandwidth of the resonance. The last four methods make an
attempt to first correct the data for rotation and translation 
(Eq. (\ref{4})), then determine $f_o$ and $Q$ of the data in canonical position.

\subsection{3 dB Method}

The 3 dB method uses the $\left| S_{21}\right| $ vs. frequency data (Fig. 
\ref{Lorentzian}), where $\left| S_{21}\right| =\sqrt{\left( 
\mathop{\rm Re}
S_{21}\right) ^2+\left( 
\mathop{\rm Im}
S_{21}\right) ^2}$. The frequency at maximum magnitude is used as the
resonant frequency, $f_o$. The half power points 
$\left( \frac 1{\sqrt{2}}\max \left| S_{21}\right| \right)$ are determined on either side of the
resonant frequency and the difference of those frequency positions is the
bandwidth $\Delta f_{3dB}$. The quality factor is then given by: 
\begin{equation}
Q=f_o/\Delta f_{3dB}  
\label{q3db}
\end{equation}
Because this method relies solely on the discrete data, not a fit, it tends
to give poor results as the signal-to-noise ratio decreases.

\subsection{Lorentzian Fit}

For this method, the $\left| S_{21}\right| $ vs. frequency data is fit to a
Lorentzian curve (Eq. (\ref{3}) and Fig. \ref{Lorentzian}) using a nonlinear
least squares fit.\cite{Bevington} The resonant frequency $f_o$, bandwidth
$\Delta f_{Lorent}$, constant background $A_1$, slope on the background 
$A_2$, skew $A_3$, and maximum magnitude $\left| S_{\max }\right| $ are used as
fitting parameters for the Lorentzian: 
\begin{equation}
\left| S_{21}\left( f\right) \right| =A_1+A_2f+\frac{\left| S_{\max }\right|
+A_3f}{\sqrt{1+4\left( \frac{f-f_o}{\Delta f_{Lorent}}\right) ^2}}
\end{equation}

The least squares fit is iterated until the change in chi squared is less
than 1 part in 10$^3$. The $Q$ is then calculated using the values of $f_o$
and $\Delta f_{Lorent}$ from the 
final fit parameters: $Q=f_o/\Delta f_{Lorent}$. 
This method is substantially more robust in the presence of
noise than the 3 dB method. For purposes of comparison with other methods,
we shall use the simple expressions for $f_o$ and $Q$ given above, rather
than the values modified by the skew parameter.

\subsection{Resonance Curve Area Method}

In an attempt to use all of the data, but to minimize the effects of noise
in the determination of $Q$, the Resonance Curve Area (RCA) method was
developed.\cite{Miura} 
In this approach the area under the 
$\left|S_{21}\left( f\right) \right| ^2$ curve is integrated to arrive at a
determination of $Q$. In detail, the RCA method uses the magnitude data
squared, $\left| S_{21}\right| ^2$, versus frequency and fits it to a
Lorentzian peak (same form as Fig. \ref{Lorentzian}): 
\begin{equation}
\left| S_{21}\left( f\right) \right| ^2=\frac{P_o}{1+4\left( \frac{f-f_o}{
\Delta f_{RCA}}\right) ^2}  
\label{10}
\end{equation}
using the resonant frequency, $f_o$, and the maximum magnitude squared, $P_o$
, as fitting parameters. The bandwidth $\Delta f_{RCA}$ is a parameter in
the Lorentzian fit, but is not allowed to vary. This method iterates the
Lorentzian fit until chi squared changes by less than 1 part in 10$^4$.
Next, using the fit values from the Lorentzian, the squared magnitude 
$\left| S_{21}\left( f_o\pm f_r\right) \right| ^2$ is found at two points 
$f_o $ $\pm $ $f_r$ on the tails of the Lorentzian far from the resonant
frequency. The area under the data, $S_1$, from $f_o$ $-$ $f_r$ to $f_o$ $+$ 
$f_r$ (symmetric positions on either side of the resonant frequency) is
found using the trapezoidal rule:\cite{Press} 
\begin{equation}
S_1=\int_{f_o-f_r}^{f_o+f_r}\left| S_{21,data}\left( f\right) \right|
^2df=\sum\limits_{N=f_o-f_r}^{f_o+f_r}\frac{\delta f}2\left( \left|
S_{21,data}\left( N\right) \right| ^2+\left| S_{21,data}\left( N+1\right)
\right| ^2\right)  
\label{integral}
\end{equation}
Here $\left| S_{21,data}\left( N\right) \right| ^2$ indicates the magnitude
squared data point at the frequency $N$, and $\delta f$ is the frequency
step between consecutive data points.

The quality factor is subsequently computed 
from the area as follows:\cite{Miura} 
\begin{equation}
Q=f_o\frac{P_o}{S_1}\tan ^{-1}\sqrt{\frac{P_o}{\left| S_{21}\left( f_o\pm
f_r\right) \right| ^2}-1}  
\label{12}
\end{equation}

This $Q$ is compared to the previously determined one. If $Q$ changes by
more than 1 part in 10$^4$, the Lorentzian fit is repeated using as initial
guesses for $f_o$ and $P_o$, the values of $f_o$ and $P_o$ from the previous
Lorentzian fit, but the fixed value of the bandwidth 
becomes $\Delta f_{RCA}=f_o/Q$. 
With the new returned parameters from the fit, $Q$ is again
computed by Eqs. (\ref{integral}) and (\ref{12}) and compared to the
previous one, and the cycle continues until convergence on $Q$ is achieved.
This method is claimed to be more robust against noise because it uses all
of the data in the integral given in Eq. (\ref{integral}).\cite{Miura}

All of the above methods assume a simple Lorentzian-like appearance of the 
$\left| S_{21}\right| $ vs. frequency data. However, the translation and
rotation of the data described by Eq. (\ref{4}) can significantly alter the
appearance of $\left| S_{21}\right| $ vs. frequency. In addition, other
nearby resonant modes can dramatically alter 
the appearance of $\left| S_{21}\right| $.\cite{Gao} 
For these reasons, it is necessary, in general,
to correct the measured $S_{21}$ data to remove the effects of cross-talk,
delay, and nearby resonant modes. The remaining methods in the section all
address these issues before attempting to calculate the $Q$ and resonant
frequency.

\subsection{Inverse Mapping Technique}

\subsubsection{Circle Fit}

The inverse mapping technique, as well as all subsequent methods in this
section, make use of the complex $S_{21}$ data and fit a circle to the plot
of $
\mathop{\rm Im}
\left( S_{21}\right) $ vs. $
\mathop{\rm Re}
\left( S_{21}\right) $ (Fig. \ref{Circles}). 
The details of fits of complex $S_{21}$ 
data to a circle have been discussed before by several 
authors.\cite{Sanchez}$^{,}$\cite{Ma} 
The data is fit to a circle using a linearized
least-squares algorithm. In the circle fit, the data is weighted by first
locating the point midway between the first and last data point; this is the
reference point $(x_{ref},y_{ref})$ (see Fig. \ref{Circles}). Next, the
distance from the reference point to each data point $(x_i,y_i)$ is
calculated. A weight is then assigned to each data point ($i$ $=$ 1 to 801)
as: 
\begin{equation}
W_{Map,i}=\left[ \left( x_{ref}-x_i\right) ^2+\left( y_{ref}-y_i\right)
^2\right] ^2  
\label{13}
\end{equation}
This gives the points closer to the resonant frequency a heavier weight than
those further away. The circle fit determines the center and radius of a
circle which is a best fit to the data.

\subsubsection{Inverse Mapping}

We now know the center and radius of the circle which has suffered
translation and rotation, as described by Eq. (\ref{4}). Rather than
un-rotating and translating the circle back into canonical position, this
method uses the angular progression of the measured points around the circle
(as seen from the center) as a function of frequency to extract the $Q$ and
resonant frequency.\cite{Taber} Three data points are selected from the
circle, one randomly chosen near the resonant frequency ($f_2$), and two
others ($f_1$ and $f_3$) randomly selected but approximately 1 bandwidth
above and below the resonant frequency (see Fig. \ref{Taberg} (b)). Figure 
\ref{Taberg} (a) shows the complex frequency plane with the measurement
frequency axis ($
\mathop{\rm Im}
f$) 
and the pole of interest at a position $if_o$ - $\Delta f_{Map}/2$. The
conformal mapping defined by: 
\begin{equation}
S_{21}=\frac{\overline{S_{21}}\Delta f_{Map}/2}{f-\left( if_o-\frac{\Delta
f_{Map}}2\right) }  
\label{14}
\end{equation}
maps the imaginary frequency axis into a circle in canonical position in the 
$S_{21}$ plane (this mapping is obtained from Eq. (\ref{1}) by rotating the
frequency plane by $e^{-i\pi /2}$). Under this transformation, a line
passing through the pole in the complex frequency plane (such as the line
connecting the pole and $if_2$ in Fig. \ref{Circles} (a)) will map into a
line of equal but opposite slope through the origin in the $S_{21}$ 
plane.\cite{Churchill} In addition, because the magnitudes of the slopes are
preserved, the angles between points $f_1$ and $f_2$ ($\theta _1$), and
points $f_2$ and $f_3$ ($\theta _2$), 
in the $S_{21}$ plane (Fig. \ref{Taberg} (b)) 
are exactly the same as those subtended from the pole in the
complex frequency plane (Fig. \ref{Taberg} (a)).\cite{mapping} The angles
subtended by these three points, as seen from the center of the circle in
the $S_{21}$ plane, define circles in the complex frequency plane which
represent the possible locations of the resonance pole (dashed circles in
Fig. \ref{Taberg} (a)).\cite{Taber}$^{,}$\cite{Richardson} The intersection
of these two circles off of the imaginary frequency axis uniquely locates
the resonance pole. The resonant frequency and $Q$ are directly calculated
from the pole position in the complex frequency plane as $f_o$ and 
$f_o/\Delta f_{Map}$. This procedure is repeated many times by again choosing
three data points as described above, and the results for $Q$ and resonant
frequency are averaged.

\subsection{Modified Inverse Mapping Technique}

We find that the fit of the complex $S_{21}$ data to a circle is critically
important for the quality of all subsequent determinations of $Q$ and $f_o$.
Hence we experimented with different ways of weighting the data to
accomplish the circle fit. The modified inverse mapping technique is
identical to the previous inverse mapping, except for a difference in the
weighting schemes for the fit of the data to a circle (Fig. \ref{Circles}).
Here the weighting on each data point, known as the standard weighting, is: 
\begin{equation}
W_{Stnd,i}=\left[ \left( x_{ref}-x_i\right) ^2+\left( y_{ref}-y_i\right)
^2\right]  
\label{15}
\end{equation}
and is the square root of the weighting in Eq. (\ref{13}). Other kinds of
weighting will be discussed in section VI.

\subsection{Phase versus Frequency Fit}

In the phase vs. frequency fit,\cite{Ma} the complex transmission data is
first fit to a circle as discussed above for the inverse mapping technique.
In addition, an estimate is made of the rotation angle of the circle. The
circle is then rotated and translated so that its center lies at the origin
of the $S_{21}$ plane (rather than canonical position), and an estimation of
the resonant frequency is found from the intersection of the circle with the
positive real axis (see Fig. \ref{Phase} inset). The phase angle of every
data point with respect to the positive real axis is then calculated. Next
the phase as a function of frequency (Fig. \ref{Phase}), obtained from the
ratio of the two parts of Eq. (\ref{5}), is fit to this form using a
nonlinear least-squares fit:\cite{Press}

\begin{equation}
\phi \left( f\right) =\phi _o+2\tan ^{-1}\left[ 2Q\left( 1-\frac f{f_o}
\right) \right]
\end{equation}
In this equation $\phi _o$, the angle at which the resonant frequency
occurs, $f_o$, and $Q$ are determined from the fit.\cite{tan} A weighting is
used in the fit to emphasize data near the resonant frequency and discount
the noisier data far from the resonance which shows little phase variation.
Again we find that the quality of this fit is sensitive to the method of
fitting the original $S_{21}$ data to a circle.

\subsection{Snortland Method}

As will be shown below, the main weakness of the Inverse Mapping and Phase
versus Frequency methods is in the initial circle fit of the complex $S_{21}$
data. To analyze the frequency dependence of the data, or to bring the
circle back into canonical position for further analysis, the center and
rotation angle (Eq. \ref{4}) must be known to very high precision. The
Snortland method makes use of internal self-consistency checks on the data
to make fine adjustments to the center and rotation angle parameters, thus
improving the accuracy of any subsequent determination of the resonant
frequency and $Q$.

The Snortland method\cite{Snortland} starts with a standard circle fit and
phase vs. frequency fit (Fig. \ref{Phase}) as discussed above. A
self-consistency check is made on the $S_{21}$ data vs. frequency by making
use of the variation of the stored energy in the resonator as the frequency
is scanned through resonance. As the resonant frequency is approached from
below, the current densities in the resonator increase. Beyond the resonant
frequency they decrease again. Hence a sweep through the resonance is
equivalent to an increase and decrease of stored energy in the cavity and
power dissipated in the sample. In general, there is a slight nonlinear
dependence of the sample resistance and inductance on resonator current $I$.
This leads to a resonant frequency and quality factor which are
current-level dependent. The generalized expression for a resonator with
current-dependent resonant frequency and $Q$ is\cite{Snortland}

\begin{equation}
s\equiv \frac{S_{21}(\omega ,I)}{S_{21}(\omega _{\max },I_{\max })}=\frac 1{
\frac{Q_{\max }}{Q(I)}+i2Q_{\max }\left( \frac{\omega -\omega _o(I)}{\omega
_o(I)}\right) }
\end{equation}
where $\omega _{\max }$ and $Q_{\max }$ are the resonant frequency and $Q$
at the point of maximum current in the resonator, $I_{\max }$. The $Q$ and
resonant frequency are therefore determined at every frequency point on the
resonance curve as\cite{Snortland}

\begin{equation}
Q(I)=\frac{Q_{\max }}{
\mathop{\rm Re}
[s^{-1}]}
\end{equation}
\begin{equation}
\omega _o(I)=\frac \omega {[1+
\mathop{\rm Im}
[s^{-1}]/2Q_{\max }]}
\end{equation}
If it is assumed that the response of the resonator is non-hysteretic as a
function of power, then the up and down ''power ramps'' must give consistent
values for the $Q$ and resonant frequency at each current level. If the data
is corrupted by a rotation in the $S_{21}$ plane, the slight nonlinear
response of $Q$ and $f_o$ with respect to field strength causes the plots of 
$Q$ and $f_o$ vs. the current level to trace out hysteresis 
curves.\cite{Snortland} By adjusting the rotation phase angle and $Q_{\max }$
parameters, one can make the two legs of the $Q(I)$ and $\omega _o(I)$
curves coincide, thereby determining the resonant frequency and $Q$ more
precisely.\cite{Snortland}

In practice, the resonant frequency is determined from a fit to the
non-linear inductance as a function of 
resonator current $I$ through $\omega(I)^{-2}=c_0+c_1I$ 
so that $\omega _{\max }=1/\sqrt{c_0+c_1I_{\max }}$. 
$Q_{\max }$ is determined by making the two legs of the $\omega _o(I)$ curve
overlap. The resulting determination of resonant frequency and quality
factor are $\omega _{\max }$ and $Q_{\max }$, respectively.

\section{Comparing Methods and Discussion}

The values of $Q$ and $f_o$ obtained by each method for a group of data
(e.g. fixed power or fixed $Q$) are averaged and their standard deviations
are determined. These results are used to compare the methods. The accuracy
of each method is determined using the generated data since, in those cases,
the true values for $Q$ and $f_o$ are known. The most accurate method is
simply the one that yields an average 
($\overline{f_o}$, $\overline{Q}$)closest to the actual value ($f_o^{known}$, $Q^{known}$). The standard
deviations ($\sigma _{f_o}$, $\sigma _Q$) for the measured data are used as
a measure of precision for the methods. The smaller the standard deviation
returned, the more precise the method. To determine the most robust method
over a wide dynamic range of $Q$ and noise, both accuracy and precision are
considered. Hence the algorithm that is both accurate and precise over
varying $Q$ or noise is deemed the most robust.

\subsection{Fixed Power Data}

Figures \ref{fop10} and \ref{Qp10} show the values of $f_o$ and $Q$
respectively, resulting from the Lorentzian fit (B), the modified inverse
mapping technique (E), and the phase vs. frequency fit (F), for the $+$10
dBm (SNR $\approx $ 108) fixed power run. For $f_o$, all three methods
return values that are very close to each other. This is verified by the
ratios of $\sigma _{f_o}/\overline{f_o}$ for those methods shown in Table 
\ref{precision table}, which shows the normalized ratio (normalized to the
lowest number) of the standard deviation of $f_o$ and $Q$ to their average 
($\sigma _{f_o}/\overline{f_o}$, $\sigma _Q/\overline{Q}$) returned by each
method on identical data. The difference in $f_o$ from trace to trace, seen
in Fig. \ref{fop10} is due entirely to the particular noise distribution on
that $S_{21}(f)$ trace. On the other hand, the determinations of $Q$ are
very different for the three methods. From Fig. \ref{Qp10}, we see that the
phase vs. frequency fit is more precise in finding $Q$ than both the
Lorentzian fit and the modified inverse mapping technique (see also Table 
\ref{precision table}). Thus the fixed power data identifies the phase vs.
frequency fit as the best.

\subsection{Power-Ramped Data}

Figures \ref{foprmp} and \ref{Qprmp} show the results for $f_o$ and $Q$
respectively, from the same methods, for the measured power-ramped data
sets. The data are plotted vs. the signal-to-noise ratio (SNR) discussed in
section III. As the SNR decreases, the determination of $f_o$ becomes less
precise, but as in the case of fixed power, all of the methods return
similar ratios for $\sigma _{f_o}/\overline{f_o}$ as confirmed 
by Table \ref{precision table}. 
The determination of $Q$ also becomes less precise as the
SNR decreases tending to overestimate its value for noisier data. But, from
Fig. \ref{Qprmp}, we see that while the modified inverse mapping technique
and phase vs. frequency fit give systematically increasing values of $Q$ as
the SNR decreases, the Lorentzian fit simply jumps around the average value.
This implies that for a low SNR, the Lorentzian fit is a more precise
method. Table \ref{precision table} confirms this statement by showing that
the Lorentzian fit has the smallest ratio of $\sigma _Q/\overline{Q}$. We
thus conclude that over a wide dynamic range of SNR the Lorentzian fit is
superior, although the phase vs. frequency fit is not significantly worse.

From figures \ref{foprmp} and \ref{Qprmp}, we see that the $f_o$
determination does not degrade nearly as much as the $Q$ determination as
SNR decreases. Here, $\sigma _{f_o}/\overline{f_o}$ changes by a factor of
2, while $\sigma _Q/\overline{Q}$ changes by a factor of 300 as SNR
decreases from 100 to 3, so the precision in the determination $f_o$ is much
greater than that of $Q$. The trend of decreasing $Q$ as the SNR increases
beyond a value of about 50 in Fig. \ref{Qprmp} is most likely due to the
non-linear resistance of the superconducting walls in the cavity. An
analysis of generated data power-ramps does not show a decreasing $Q$ at
high SNR.

\subsection{Precision, Accuracy, and Robustness}

The most precise methods over different fixed powers are the nonlinear least
squares fit to the phase vs. frequency (F) and the Lorentzian nonlinear
least squares fit (B) (Table \ref{precision table}). They consistently give
the smallest ratios of their standard deviation to their average for both $Q$
and $f_o$ compared to all other methods. At high power (SNR 
\mbox{$>$}
350) the phase vs. frequency fit is precise to about 3 parts in 10$^{10}$
for the resonant frequency and to 3 parts in 10$^4$ for the quality factor,
when averaged over about 75 traces.

When looking at the generated data with SNR $\approx $ 65, the most accurate
method for the determination of the resonant frequency is the phase vs.
frequency fit, because it returns an average closest to the true value, or
as in Table \ref{accuracy table}, it has the smallest ratio of the
difference between the average and the known value divided by the known
value ($\frac{\left| \overline{f_o}-f_o^{known}\right| }{f_o^{known}}$, 
$\frac{\left| \overline{Q}-Q^{known}\right| }{Q^{known}}$). The value
returned for the resonant frequency is accurate to about 8 parts in 10$^8$
for $Q$ = 10$^3$, and 1 part in 10$^9$ for $Q$ = 10$^5$ when averaged over
100 traces. For the quality factor, the phase vs. frequency fit (F) is most
accurate (Table \ref{accuracy table}), with accuracy to about 1 part in 
10$^4 $ for $Q$ = 10$^3$, and 1 part in 10$^4$ for $Q$ = 10$^5$ when averaged
over 100 traces.

The method most robust in noise is the Lorentzian fit (see the power ramp
columns of both Tables). It provided values for $f_o$ and $Q$ that were the
most precise and accurate as the signal-to-noise ratio decreased
(particularly for SNR 
\mbox{$<$}
10). Over several decades of $Q$, the most robust method for the
determination of $f_o$ is the phase vs. frequency fit, which is precise to
about 1 part in 10$^5$ when $Q$ = 10$^2$, and to about 1 part in 10$^8$ when 
$Q$ = 10$^5$, averaged over 100 traces with SNR $\approx $ 65. For the
determination of $Q$, the phase vs. frequency (F) is also the most robust,
providing precision to 2 parts in 10$^3$ when $Q$ = 10$^2$ to 10$^5$
averaged over 100 traces.

\section{Improvements}

The first three methods discussed above (3dB, Lorentzian fit, and RCA
method) can be improved by correcting the data for rotation and translation
in the complex $S_{21}$ plane. All of the remaining methods can be improved
by carefully examining the validity of the circle fit. We have observed that
by modifying the weighting we can improve the fit to the circle for noisy
data, and thereby improve the determination of $Q$ and $f_o$. For instance,
Fig. \ref{weighting} shows that the standard weighting (the weighting from
the modified inverse mapping technique) systematically overestimates the
radius of the circle for noisy data. Below we discuss several ways to
improve these fits.

By introducing a radial weighting, we can improve the circle fit
substantially (an example is shown in Fig. \ref{weighting}). For the radial
weighting, we first do the standard weighting to extract an estimate for the
center of the circle $(x_c,y_c)$, which is not strongly corrupted by noise.
The radial weighting on each point ($i=$ 1 to 801) is then defined as: 
\begin{equation}
W_{Radial,i}=\frac 1{\sqrt{\left( x_c-x_i\right) ^2+\left( y_c-y_i\right) ^2}
}  
\label{17}
\end{equation}
which reduces the influence of noisy data points well outside the circle.
Figure \ref{radwsnr} shows a plot of the calculated radius versus the
signal-to-noise ratio for the generated power-ramped data set. The figure
shows plots of the calculated radius using four different weightings: 
$W_{stnd}$ (Eq. (\ref{15})), $W_{Radial}$ (Eq. (\ref{17})), 
$W_{Radial}^{1/2}$, and $W_{Radial}^2$. From this plot, it is clear that above a SNR of about
30 all of the weightings give very similar radius values. However, below
that value we see that the radius from the $W_{Radial}^{1/2}$ weighting
agrees best with the true radius of 0.2. Therefore, by improving the circle
fit with a similar weighting scheme, we hope to extract even higher
precision and better accuracy from these methods at lower signal-to-noise
ratio.

In addition to errors in the fit radius of the circle at low SNR, there can
also be errors in the fit center of the circle. Figure \ref{centerror} shows
the normalized error, $E_c$: 
\begin{equation}
E_c=\sqrt{\left( \frac{x_c-x_{fit}}{x_c}\right) ^2+\left( \frac{y_c-y_{fit}}{
y_c}\right) ^2}
\end{equation}
in the calculation of the center of the circle from weightings: $W_{stnd}$, 
$W_{Radial}$, $W_{Radial}^{1/2}$, and $W_{Radial}^2$, vs. the SNR in log
scaling. Here $\left( x_c,y_c\right) $ is the true center of the circle and 
$\left( x_{fit},y_{fit}\right) $ is the calculated center from the circle
fit. From Fig. \ref{centerror}, we see that the calculation of the center of
the circle is accurate to within 1\% for SNR $\approx $ 20 and above using
any weighting. However, below SNR = 10, all of the weightings give degraded
fits. The inset (b) of Fig. \ref{centerror} shows the angle $\alpha $ vs.
SNR, where $\alpha $ is the angle between the vector connecting the true and
calculated centers, and the vector connecting the true center to the
position of the resonant frequency. From this figure we see that the angle
between these vectors approaches $\pi $ as SNR decreases, which means that
the fit center migrates in the direction away from the resonant frequency as
the data becomes noisy. This indicates that the points on the side of the
circle opposite from the resonant frequency have a combined weight larger
than those points around the resonant frequency, and thus the center is
calculated closer to those points.

For data with SNR greater that about 10, all weightings give similar results
for the circle fits. For data with SNR less than 10, the best circle fit
would make an estimate of the radius of the circle by using the square root
radial weighting, and an estimate of the center by weighting data near the
resonant frequency more heavily.

A further refinement of the inverse mapping method would be to fit the data
with an arbitrary number of poles and zeroes to take account of multiple
resonances in the frequency spectrum.\cite{Richardson}

The Snortland method was originally developed to analyze non-linear
resonances.\cite{Snortland} Our use of it for linear low-power resonances
was preliminary, and the results probably do not reflect its ultimate
performance. Further development of this method on linear resonances has the
potential to produce results superior to those obtained with the phase vs.
frequency method at high SNR.

\section{Conclusions}

We find that the phase versus frequency fit and the Lorentzian nonlinear
least squares fit are the most reliable procedures for estimating $f_o$ and 
$Q$ from complex transmission data. The Lorentzian fit 
of $\left|S_{21}\right| $ vs. 
frequency is surprisingly precise, but suffers from poor
accuracy relative to vector methods, except for very noisy data. However, a
major advantage of vector data is that it allows one to perform corrections
to remove cross talk, delay, and nearby resonances, thus significantly
improving the quality of subsequent fits. For the fixed-power measured data
sets, the phase vs. frequency fit has the highest precision and accuracy in
the determination of $f_o$ and $Q$ making it the best method overall. All of
these methods are good for SNR greater than about 10. Below this value, all
methods of determining $Q$ and resonant frequency from complex transmission
coefficient data degrade dramatically. Concerning robustness, the phase vs.
frequency fit does well for a dynamic range of $Q$, while the Lorentzian fit
does well in the power-ramp (SNR = 1 to 2000).

We also find that significant improvements can be made to the determination
of resonant frequency and $Q$ in noisy situations when careful attention is
paid to the circle fitting procedure of the complex $S_{21}$ data. Further
development of the inverse mapping and Snortland methods can greatly improve
the accuracy and precision of resonant frequency and Q determination in
realistic measurement situations.

\acknowledgements We thank A. Schwartz and B. J. Feenstra for their critical
reading of the manuscript, and H. J. Snortland and R. C. Taber for many
enlightening discussions. This work is supported by the National Science
Foundation through grant number DMR-9624021, and the Maryland Center for
Superconductivity Research.

\squeezetable

\begin{table}[tbp] \centering
\caption{Measurements of relative precision of 
the seven methods used to determine 
$f_o$ and $Q$ from complex transmission data.  
Tabulated are ratios of the standard 
deviation to the average values for both resonant 
frequency ($\sigma(f_o)/\overline{f_o}$) and 
quality factor ($\sigma(Q)/\overline{Q}$) 
normalized to the best value (given in parentheses), 
for SNR $\approx $ 49, 368, 
and ramped from 5 to 168. All entries are based on measured data.
\label{precision table}} 
\begin{tabular}{lllllll}
Precision Table & \multicolumn{2}{l}{Noisy (P = +3 dBm, SNR $\approx $ 49)}
& \multicolumn{2}{l}{Less Noisy (P = +15 dBm, SNR $\approx $ 368)} & 
\multicolumn{2}{l}{Power Ramp (SNR $\approx $ 5 to 168 )} \\ 
Method & $Q$ & $f_o$ & $Q$ & $f_o$ & $Q$ & $f_o$ \\ 
3 dB & 5.91 & 1.069 & 7.50 & 4.77 & 190.44 & 1.274 \\ 
Lorentzian & 1.55 & 1.025 & 2.27 & 1.10 & 1 (1.91 x 10$^{-2}$) & 1.004 \\ 
RCA & 5.66 & 1.030 & 5.24 & 1 & 11.04 & 1.031 \\ 
Inverse Mapping & 6.02 & 1.021 & 7.95 & 1.57 & 4.27 & 1.321 \\ 
Modified Mapping & 1.49 & 1.031 & 5.89 & 2.13 & 1.61 & 1 (7.17 x 10$^{-9}$)
\\ 
Phase vs. Freq & 1 (2.51 x 10$^{-3}$) & 1 (1.15 x 10$^{-8}$) & 1 (2.80 x 10$
^{-4}$) & 1 (3.12 x 10$^{-10}$) & 1.47 & 1.025 \\ 
Snortland & 2.27 & 1.029 & 2.09 & 1 & 5.98 & 1.086
\end{tabular}
\end{table}
\newpage
\begin{table}[tbp] \centering
\caption{Measurements of the relative accuracy of the seven methods used to 
determine $f_o$ and $Q$ from complex transmission data.  Tabulated are ratios 
of the difference of the averages of $f_o$ and $Q$ from the 
known value divided by the known values, for both 
resonant frequency 
($\frac{\left| \overline{f_o}-f_o^{known}\right| }{f_o^{known}}$)
and quality factor 
($\frac{\left| \overline{Q}-Q^{known}\right| }{Q^{known}}$). The
entries are normalized to the best value (given in parentheses), 
for $Q$ = 10$^3$, 
$Q$ = 10$^5$ (both with SNR $\approx $ 65), and SNR ramped from 1 to 2000.
All entries are based on generated data.
\label{accuracy table}} 
\begin{tabular}{lllllll}
Accuracy Table & \multicolumn{2}{l}{$Q$ = 10$^3$} & \multicolumn{2}{l}{$Q$ =
10$^5$} & \multicolumn{2}{l}{Power Ramp (SNR $\approx $ 1 to 2000 )} \\ 
Method & $Q$ & $f_o$ & $Q$ & $f_o$ & $Q$ & $f_o$ \\ 
3 dB & 253.08 & 217.39 & 240.21 & 117.15 & 401.48 & 43.87 \\ 
Lorentzian & 15.38 & 27.25 & 14.93 & 17.28 & 1 (3.11 x 10$^{-2}$) & 1 (1.46
x 10$^{-9}$) \\ 
RCA & 246.15 & 403.05 & 23.35 & 217.76 & 8.39 & 73.39 \\ 
Inverse Mapping & 3.85 & 3.01 & 10.43 & 2.21 & 2.84 & 5.72 \\ 
Modified Mapping & 2.77 & 3.5 & 5.64 & 1.57 & 1.83 & 8.43 \\ 
Phase vs. Freq & 1 (1.30 x 10$^{-4}$) & 1 (7.88 x 10$^{-8}$) & 1 (1.40 x 10$
^{-4}$) & 1 (1.46 x 10$^{-9}$) & 4.03 & 12.00 \\ 
Snortland & 103.08 & 12.68 & 95.21 & 8.50 & 5.11 & 13.50
\end{tabular}
\end{table}

\begin{figure}[tbp]
\caption{Measured magnitude of the complex transmission coefficient $S_{21}$
of a superconducting resonator as a function of frequency for measured data
(Input power = $+$10 dBm, SNR $\approx $ 108). A Lorentzian curve is fit to
the data as described in the text. Inset is the lumped element model circuit
diagram for the resonator. The input and output transmission lines have
impedance $Z_o$, $l_{m1}$ and $l_{m2}$ are coupling mutual inductances, $C$
is the capacitance, $R$ is the resistance, and $L$ is the inductance of the
model resonator.}
\label{Lorentzian}
\end{figure}

\begin{figure}[tbp]
\caption{Measured imaginary vs. real part of the complex transmission
coefficient $S_{21}$ for a single resonant mode (Input power = $+$3 dBm, SNR 
$\approx $ 49). This plot shows data and a circle fit, as well as the
translated and rotated circle in canonical position. ($X$ $\approx $ ($1.67$ 
$\times$ $10^{-4}$, $-2.52$ $\times$ $10^{-4}$), $\phi$ $\approx $ $116^o$).
Large dots indicate centers of circles, and the size of the translation
vector has been exaggerated for clarity.}
\label{Circles}
\end{figure}

\begin{figure}[tbp]
\caption{(a). The complex frequency plane is shown with frequency points 
$f_1 $, $f_2$, and $f_3$ on the imaginary axis and a pole off of the axis.
The imaginary frequency axis is mapped onto the complex $S_{21}$ plane (b)
as a circle in canonical position, and the corresponding frequency points
are indicated on the circumference of the circle.}
\label{Taberg}
\end{figure}

\begin{figure}[tbp]
\caption{Measured phase as a function of frequency for measured data 
(SNR $\approx $ 31), both data and fit are shown. Inset is the translated and
rotated circle, where its center is at the origin and the phase to each
point is calculated from the positive real axis.}
\label{Phase}
\end{figure}

\begin{figure}[tbp]
\caption{Plot of fit resonant frequency versus trace number for measured
data when the source power is +10 dBm. Results are shown for three methods
discussed in the text.}
\label{fop10}
\end{figure}

\begin{figure}[tbp]
\caption{Plot of fit quality factor versus trace number for measured data
when the power is +10 dBm. Results are shown for three methods discussed in
the text.}
\label{Qp10}
\end{figure}

\begin{figure}[tbp]
\caption{Plot of fit resonant frequency versus the signal-to-noise ratio on
a log scale for the measured power-ramped data set. Results are shown for
three methods discussed in the text.}
\label{foprmp}
\end{figure}

\begin{figure}[tbp]
\caption{Plot of fit quality factor versus the signal-to-noise ratio on a
log scale for the measured power-ramped data set. Results are shown for
three methods discussed in the text.}
\label{Qprmp}
\end{figure}

\begin{figure}[tbp]
\caption{Measured imaginary vs. real part of the complex transmission
coefficient for measured data 
(SNR $\approx $ 4, $X$ $\approx $ ($7.22$ $\times$ $10^{-5}$, $3.26$ $\times$ $10^{-4}$), $\phi$ $\approx $ $220^o$).
Plot shows data and two circle fits, one where the standard weighting is
used (dashed line), and one where the square root radial weighting is used
(solid line).}
\label{weighting}
\end{figure}

\begin{figure}[tbp]
\caption{The calculated circle fit radius vs. the signal-to-noise ratio on a
log scale is shown for the generated power-ramped data set. The plot shows
the results from four different weightings: 
$W_{Stnd}$, $W_{Radial}$, $W_{Radial}^2$, $W_{Radial}^{1/2}$. The true value for the radius is 0.2.}
\label{radwsnr}
\end{figure}

\begin{figure}[tbp]
\caption{The normalized error in the determination of the center of the fit
circle is shown vs. the signal-to-noise ratio on a log scale for the
generated power-ramp data. Results are from the weightings: $W_{Stnd}$, 
$W_{Radial}$, $W_{Radial}^2$, $W_{Radial}^{1/2}$. Inset (a) is a plot of the
true circle (solid line) and the fit circle to the data (dashed line) to
show that the determination of the center from the fit $(x_{fit},y_{fit})$
is located at an angle $\alpha$ from the line connecting the true center 
$(x_c,y_c)$ to the resonant frequency point, $f_o$. The distance from the
true center to the calculated center is related to the normalized error in
the calculated center $E_c$. Inset (b) is a plot of the angle $\alpha$ vs.
log of SNR for the generated data.}
\label{centerror}
\end{figure}

\begin{references}
\bibitem{Montgomery}  C. G. Montgomery, Technique of Microwave Measurements,
MIT Rad. Lab. Series, Vol. 11 (McGraw-Hill, Inc. 1947).
\bibitem{Malter}  L. Malter, and G. R. Brewer, J. Appl. Phys., {\bf 20}, 918
(1949).
\bibitem{Ginzton}  E. L. Ginzton, {\it Microwave Measurements}, (McGraw-Hill
Inc. 1957).
\bibitem{Sucher}  M. Sucher, and J. Fox, {\it Handbook of Microwave
Measurements}, Vol. II, 3rd ed. (John Wiley \& Sons, Inc. 1963), Chap. 8.
\bibitem{Kajfez}  D. Kajfez, and E. J. Hwan, IEEE Trans. Microwave Theory
Tech., {\bf MTT-32}, 666 (1984).
\bibitem{Sun}  E. Sun, and S. Chao, IEEE Trans. Microwave Theory Tech. 
{\bf 43}, 1983 (1995).
\bibitem{Barmatz}  M. B. Barmatz, NASA Tech Brief, {\bf 19}, No. 12, Item
\#16, (Dec. 1995).
\bibitem{Padamsee}  H. Padamsee, J. Knobloch, and T. Hays, {\it RF
Superconductivity for Accelerators}, (John Wiley \& Sons, Inc. 1998).
\bibitem{Harvey}  A. F. Harvey, {\it Microwave Engineering}, (Academic Press
1963).
\bibitem{Stein}  S. R. Stein, and J. P. Turneaure, Electronics Letters, 
{\bf 8}, 321 (1972).
\bibitem{Klein}  O. Klein, S. Donovan, M. Dressel, and G. Gr\"{u}ner,
International Journal of Infrared and Millimeter Waves, {\bf 14}, 2423
(1993);S. Donovan, O. Klein, M. Dressel, K. Holczer, and G. Gr\"{u}ner,
International Journal of Infrared and Millimeter Waves, {\bf 14}, 2459
(1993);M. Dressel, O. Klein, S. Donovan, and G. Gr\"{u}ner, International
Journal of Infrared and Millimeter Waves, {\bf 14}, 2489 (1993)
\bibitem{Luiten}  A. N. Luiten, A. G. Mann, and D. G. Blair, Meas. Sci.
Technol., {\bf 7}, 949 (1996).
\bibitem{Aitker}  J. E. Aitken, Proc. IEE, {\bf 123}, 855 (1976).
\bibitem{Ashley}  J. R. Ashley, and F. M. Palka, The Microwave Journal, 35
(June 1971).
\bibitem{Watanabe}  K. Watanabe, and I. Takao, Rev. Sci. Instrum., {\bf 44},
1625 (1973).
\bibitem{Subramanian}  V. Subramanian, and J. Sobhanadri, Rev. Sci.
Instrum., {\bf 65}, 453 (1994).
\bibitem{Sanchez}  M. C. Sanchez, E. Martin, J. M. Zamarro, IEE Proceedings, 
{\bf 136}, 147 (1989).
\bibitem{Moser}  E. K. Moser, and K. Naishadham, IEEE Trans. on Appl.
Superconductivity, {\bf 7}, 2018 (1997).
\bibitem{Ma}  Z. Ma, Ph. D. Thesis, (Ginzton Labs Report No. 5298), Stanford
University, (1995).
\bibitem{Leong}  K. Leong, J. Mazierska, and J. Krupka, 1997 IEEE MTT-S
Proceedings.
\bibitem{Snortland}  H. J. Snortland, Ph. D. Thesis, (Ginzton Labs Report
No. 5552), Stanford University, (1997). See also
http://loki.stanford.edu/Vger.html. We used V'Ger version 2.37 for the
analysis performed in this paper.
\bibitem{Mao}  J. Mao, Ph. D. Thesis, University of Maryland, (1995).
\bibitem{Sridhar}  S. Sridhar, and W. L. Kennedy, Rev. Sci. Instrum., 
{\bf 59}, 531 (1988).
\bibitem{Press}  W. H. Press, B. P. Flannery, S. A. Teukolsky, W. T.
Vetterling, {\it Numerical Recipes}, (Cambridge University Press 1989), pp.
105, 202-203, 523-528.
\bibitem{Bevington}  P. R. Bevington, {\it Data Reduction and Error Analysis
for the Physical Sciences}, (McGraw-Hill, Inc. 1969), pp. 237-240.
\bibitem{Miura}  T. Miura, T. Takahashi, and M. Kobayashi, IEICE Trans.
Electron., {\bf E77-C}, 900 (1994).
\bibitem{Gao}  F. Gao, M. V. Klein, J. Kruse, and M. Feng, IEEE Trans.
Microwave Theory Tech., {\bf 44}, 944 (1996).
\bibitem{Taber}  R. C. Taber, Hewlett-Packard Laboratories, private
communication.
\bibitem{Churchill}  R. V. Churchill, and J. W. Brown, {\it Complex
Variables and Applications}, 5th Ed. (McGraw-Hill, Inc., New York) p. 209.
\bibitem{mapping}  The mapping defined by Eq. (\ref{14}) maps lines of slope 
$m$ through the pole in the frequency plane to lines of slope $-m$ through
the origin in the $S_{21}$ plane. Note the mapping defined by Eq (\ref{14})
is not conformal at the pole because it is not analytic there, however it is
an isogonal mapping at that point. See reference 29.
\bibitem{Richardson}  M. H. Richardson and D. L. Formenti, Proc.
International Modal Analysis Conference, 167 (1982).
\bibitem{tan}  The factor of 2 in front of the $\tan ^{-1}$ comes from the
difference in angle subtended as seen from the origin vs. the center of the
circle for a circle in canonical position in the $S_{21}$ plane 
(Fig. \protect{\ref{Taberg}} (b)).
\end{references}
\end{document}